\documentclass[a4paper]{article}

\usepackage{INTERSPEECH2019}
\usepackage{hyperref}
\usepackage{catchfilebetweentags} 
\usepackage{amsfonts} 
\usepackage{mathtools} 
\usepackage{amsthm} 
\usepackage{multicol}
\usepackage{subcaption}
\usepackage{soul} 
\usepackage{tabularx}
\usepackage{array}
\usepackage{xcolor}
\usepackage{thmtools, thm-restate} 

\usepackage{caption}
\newcommand{\embed}[1]{\text{embed}\left( #1 \right)}
\newcommand{\loss}{\ell}

\newcommand{\mixingfunc}{\phi}
\newcommand{\hyperparam}{\alpha}
\newcommand{\lossmixup}{\textit{loss mixup }}

\newcommand{\learnablelossmixup}{\textit{learnable loss mixup }}
\newcommand{\Learnablelossmixup}{\textit{Learnable loss mixup }}
\newcommand{\network}{f}
\newcommand{\dataset}{\mathcal{D}}
\newcommand{\f}[1]{f\left( #1 \right)}
\newcommand{\mixupparam}{\lambda}
\newcommand{\mixup}[1]{\tilde{#1}}
\newcommand{\sample}[2]{#1_{#2}}
\newcommand{\noisy}{\boldsymbol{x}}
\newcommand{\speech}{\boldsymbol{s}}
\newcommand{\noise}{\boldsymbol{n}}
\newcommand{\nsamples}{M}

\newcommand{\loadtable}[1]{%
    \ExecuteMetaData[tex/tables.tex]{#1}%
}

\newcommand{\loadfig}[1]{%
    \ExecuteMetaData[tex/figures.tex]{#1}%
}

\newcommand{\loadeq}[1]{%
    \ExecuteMetaData[tex/equations.tex]{#1}%
}

\newcommand{\loadthm}[1]{%
    \ExecuteMetaData[tex/theorems.tex]{#1}%
}

\newtheorem{definition}{Definition}[section]

\title{Single-channel speech enhancement using learnable loss mixup}
\name{Oscar Chang$^1$$^*$, Dung N. Tran$^2$, Kazuhito Koishida$^2$ \thanks{$^*$ work done during internship at Microsoft.}}
\address{$^1$Columbia University, USA\\ $^2$Microsoft Corporation, USA}
\email{oscar.chang@columbia.edu, \{dung.tran, kazukoi\}@microsoft.com}

\begin{document}

\maketitle
\begin{abstract}

Generalization remains a major problem in supervised learning of single-channel speech enhancement. In this work, we propose \textit{learnable loss mixup (LLM)}, a simple and effortless training diagram, to improve the generalization of deep learning-based speech enhancement models. \textit{Loss mixup}, of which \textit{learnable loss mixup} is a special variant, optimizes a mixture of the loss functions of random sample pairs to train a model on virtual training data constructed from these pairs of samples. In \textit{learnable loss mixup}, by conditioning on the mixed data, the loss functions are mixed using a non-linear mixing function automatically learned via neural parameterization. Our experimental results on the VCTK benchmark show that \learnablelossmixup achieves 3.26 PESQ, outperforming the state-of-the-art.

\end{abstract}

\section{Introduction}

Speech enhancement, the problem of extracting a clean speech signal from its noisy version, plays a crucial role in speech applications and has been extensively investigated in the literature. Classical approaches employ signal processing algorithms \cite{Boll1979, Lim1978, Lim1979, Malah1985} and produce satisfactory results in stationary noise environments. However, they fail in more challenging scenarios, such as nonstationary noise, and have been surpassed by deep supervised learning methods \cite{Pascual2017, Williamson2017, Rethage2018, Germain2018, Soni2018, Tan2018, Bulut2020}. Deep supervised learning for speech enhancement uses pairs of clean and noisy speech signals to train an expressive deep neural network (DNN) able to robustly predict clean speech in highly non-stationary noise environments. They have become the workhorse for modern speech enhancement systems.

State-of-the-art supervised learning-based speech enhancement in the literature typically trains the model by solving an \textit{Empirical Risk Minimization (ERM)} problem \cite{Mohri2012}. This approach, however, has a major drawback of learning the network behavior only on the training samples. This limitation leads to undesirable behavior of the model beyond the training set. A straightforward and common remedy is to extend the training set which, unfortunately, is expensive and laborious to acquire and label in speech applications. Another common approach is data augmentation which, however, is domain-dependent and thus requires expert knowledge. Improving generalization, the network's ability to produce robust predictions in unseen noise environments, thus remains a major challenge in speech enhancement.

\textbf{Our contribution.} To tackle this problem, we introduce \textit{loss mixup (LM)} with a special variant called \textit{learnable loss mixup (LLM)}. During training, \lossmixup constructs virtual samples by mixing random noisy training samples. It, then, trains the network on the augmented data by minimizing a mixture of the loss functions of the individual mixing samples. In essence, \lossmixup is not only a data augmentation method but also an instance of Vicinal Risk Minimization (VRM) \cite{chapelle2000}, a training regime optimizing the loss function within the vicinity of the training set, suitable for regression problems such as speech enhancement. The advantages of \lossmixup are threefold. First, it augments the training data with richer noise and interference conditions via virtual noisy samples. This data augmentation procedure is domain-independent and can be implemented in a few lines of code without expert knowledge. Moreover, by monitoring the network in the vicinity of training samples, \lossmixup encourages smooth regression curves and thus stabilizes the network behavior outside the training set. Finally, optimizing the loss mixture instead of the target mixture retains the individual clean targets of the mixing noisy samples, avoiding the degradation of the network in regression settings \cite{tanaka2018joint}.

Manually tuning the mixing parameter, sampled from a mixing distribution in our framework, for the loss mixture is nontrivial. Therefore, we propose \textit{learnable loss mixup} to automatically learn the mixing distribution via gradient descent by reparameterizing the mixing distribution parameters into loss function variables. Furthermore, we introduce conditions allowing reparameterization by neural networks which can be trained efficiently via backpropagation. On the VCTK speech enhancement dataset \cite{valentini2016speech}, we show that \learnablelossmixup significantly outperforms standard ERM training. It achieves 3.26 PESQ, surpassing the previous state-of-the-art by 0.06 points.

\textbf{Related work.} Our method resembles the sample mixing mechanism of \textit{mixup} \cite{zhang2017mixup}, a new training method aiming to improve the generalization of supervised learning for classification problems. Similar to our \textit{loss mixup}, mixup, which we call \textit{label mixup} in this paper,  adopts the Vicinal Risk Minimization approach by generating virtual samples from random training samples. However, unlike \textit{loss mixup}, label mixup trains the network on virtual targets obtained from the labels of the respective mixing samples. As label mixing is a form of label smoothing \cite{szegedy2016rethinking}, label mixup is beneficial to classification tasks and has achieved both better generalization and increased model calibration in classification applications across various data domains such as images, audio, text, and tabular data \cite{zhang2017mixup,thulasidasan2019mixup}. Nevertheless, applying label mixup to regression problems, such as speech enhancement, is inappropriate since mixing clean targets produces noisy labels which degrade the network performance in regression settings \cite{tanaka2018joint}. As mentioned above, \lossmixup overcomes this issue by retaining the targets of the mixing samples and optimizing the mixture of the respective individual loss functions. This insight is supported by our ablation study, in the experiment section, confirming the superiority of \lossmixup over \textit{label mixup} in speech enhancement. To our knowledge, we are the first to propose the sample mixup principle beyond non-classification tasks.

\textbf{Outline.} The rest of the paper is organized as follows. We introduce \lossmixup in Section~\ref{sec:lossmixup} and draw connection between \lossmixup and \textit{label mixup}. In Section~\ref{sec:learnablelossmixup}, we present \learnablelossmixup by detailing the reparameterization trick and conditions allowing an efficient neural parameterization of the mixing function. We discuss experimental results and their implications in Section~\ref{sec:experiment}, and, finally, conclude our work in Section~\ref{sec:conclusions}.

\section{Loss mixup}\label{sec:lossmixup}
We seek an enhanced version of an observed noisy speech signal $\noisy$ via a learning-based approach. As commonly the case in practice, the noisy speech is assumed to follow an additive noise model:
    \loadeq{eq:noisymodel}
where $\speech$ and $\noise$ are the unknown clean speech and the unknown additive noise, respectively. We train a deep neural network (DNN) $\network$ to extract the clean speech from the noisy speech without noise information.

We adopt a supervised learning regime that trains the DNN on a dataset $\dataset = \{(\noisy_j, \speech_j)\}_j^\nsamples$ of 
$\nsamples$ pairs of noisy and clean speech. Let $\loss{}$ be a loss function enforcing the consistency between the enhanced speech and the corresponding clean speech, current state-of-the-art methods typically train the network by solving the \textit{empirical risk minimization (ERM)} problem:
    \loadeq{eq:erm}
where $\f{\sample{\noisy}{j}}$ is the prediction for noisy sample $\sample{\noisy}{j}$. Despite being the workhorse of modern speech enhancement systems, ERM possesses a major drawback: It causes the network to memorize the training data by limiting the training on a finite set of samples. Data memorization, in turn, leads to poor network generalization.

\subsection{Loss mixup training}
To alleviate the generalization issue, we propose \textit{loss mixup (LM)} training, a generalized version of \textit{mixup} \cite{zhang2017mixup}, for speech enhancement. Given two random training samples $(\sample{\noisy}{j}, \sample{\speech}{j})$ and $(\sample{\noisy}{k}, \sample{\speech}{k})$, and a mixing parameter $\mixupparam \in [0, 1]$, \lossmixup constructs virtual noisy samples
    \loadeq{eq:mixup-sample}
and trains the network by minimizing 
    \loadeq{eq:loss-mixup-loss}
Importantly, the virtual noisy samples are constructed during the training stage only. During inference, for a noisy speech signal $\noisy$, the enhanced speech is obtained from the network's output $\network(\noisy)$.

Similar to label mixup, the success of \lossmixup is highly sensitive to the mixing parameter $\mixupparam$ \cite{guo2019mixup,thulasidasan2019mixup,verma2019manifold}. For instance, Zhang et. al. \cite{zhang2017mixup} proposed for $\lambda$ to be drawn from a symmetric Beta distribution $\beta(\alpha,\alpha)$ which varies dramatically for different values of $\alpha$ as demonstrated in Fig.~\ref{fig:beta_visualization}. Moreover, several studies \cite{zhang2017mixup, guo2019mixup} indicate that the optimal mixing distribution is dataset dependent, and manually tuning the mixing distribution to fit a given dataset is expensive. This challenge motivates our \learnablelossmixup variant in Section~\ref{sec:learnablelossmixup}. 

    \loadfig{beta-visualization}
    
\subsection{Connection to label mixup}
    Before introducing \learnablelossmixup in the next section, here, we draw a connection between \lossmixup and label mixup and highlight that applying them to speech enhancement yields different model performance.
    \loadthm{thm:label-v-loss}
    \loadthm{proof:label-v-loss}
    \loadthm{corr:mixup-classification}

    Corollary~\ref{eqn:mixup_classification} implies that \lossmixup and label mixup are equivalent for classification and regression tasks utilizing the cross-entropy and mean squared error loss. However, for application-specific loss functions such as the log-spectral distance (LSD), a common loss function in speech applications, \lossmixup and label mixup lead to different model performance for speech enhancement. Our ablation study in Section~\ref{sec:experiment} confirms this insight.



\section{Learnable loss mixup}\label{sec:learnablelossmixup}

\subsection{The reparameterization trick}
We follow the approach in \cite{zhang2017mixup} that samples $\mixupparam$ from a mixing distribution $p(\hyperparam)$. As discussed in the previous section, manually tuning the mixing distribution is impractical. Furthermore, directly learning the hyperparameters $\hyperparam$ of the mixing distribution is unfeasible since they are absent from the loss function. Therefore, we reparameterize the mixed loss function (\ref{eq:loss-mixup-loss}) by sampling $\mixupparam$ from the uniform distribution $\mathcal{U}(0,1)$ and shaping $\mixupparam$ with a parameterized mixing function $\mixingfunc$:
    \loadeq{eq:learnable-loss-mixup-loss}
    
\Learnablelossmixup (LLM), which trains the network by minimizing (\ref{eq:learnable-loss-mixup-loss}) instead of (\ref{eq:loss-mixup-loss}), allows using gradient descent to learn the hyperparameters of the mixing distribution. In particular, assume $\mixupparam$ follows an arbitrary distribution $p(\hyperparam)$ and let $\mixingfunc_\hyperparam$ be the inverse cumulative distribution function (CDF) of $p(\hyperparam)$, the reparameterization trick enables us to calculate Monte Carlo estimates of the loss function differentiable with respect to (w.r.t.) $\hyperparam$:

\begin{equation}
\label{eqn:reparam_trick}
\begin{split}
\nabla_\alpha \mathbb{E}_{\lambda \sim p(\hyperparam)} \mathcal{L}(\lambda) &= \nabla_\alpha \mathbb{E}_{\lambda \sim \mathcal{U}(0,1)}\mathcal{L}(\phi_\alpha(\lambda))\\
&= \mathbb{E}_{\lambda \sim \mathcal{U}(0,1)} \nabla_\alpha \mathcal{L}(\phi_\alpha(\lambda)).\\
\end{split}
\end{equation}
The optimal mixing distribution $p(\hyperparam)$, then, is obtained by minimizing the last term of (\ref{eqn:reparam_trick}) w.r.t. $\hyperparam$ via gradient descent.

Eq.~(\ref{eqn:reparam_trick}) implies that any inverse CDF is a candidate for the reparametrization of a mixing distribution. This insight motivates us to define below desired properties of the mixing function $\phi$. They allow $\mixingfunc$ to be parameterized by neural networks which can be trained efficiently via backpropagation.

    \subsection{Representation theorem for the mixing function}
    
    \begin{definition}
\label{eqn:mixing_fn}
$\phi: [0,1] \rightarrow [0,1]$ is a \emph{mixing function} if 
\begin{enumerate}
    \item $\phi(1) = 1$,
    \item $\forall \lambda \in [0,1]: 1-\phi(\lambda) = \phi(1-\lambda),$
    \item and $\phi$ is monotonically increasing.
\end{enumerate}
\end{definition}
The first part of the definition implies \lossmixup reverts to ERM when $\lambda=1$. The second part enforces symmetry: \lossmixup produces the same loss when $(\sample{\noisy}{j}, \sample{\speech}{j})$ and $(\sample{\noisy}{i}, \sample{\speech}{i})$ are switched and $\lambda$ becomes $1-\lambda$. The third part ensures that $\phi$ is an inverse CDF. Our definition implies a general representation for mixing functions that reduces the parametrization of $\phi$ to a simpler function $\rho$.

    \loadthm{thm:representation}

    \loadthm{proof:representation}
    
\begin{figure}[t]
\centering
\includegraphics[width=\linewidth]{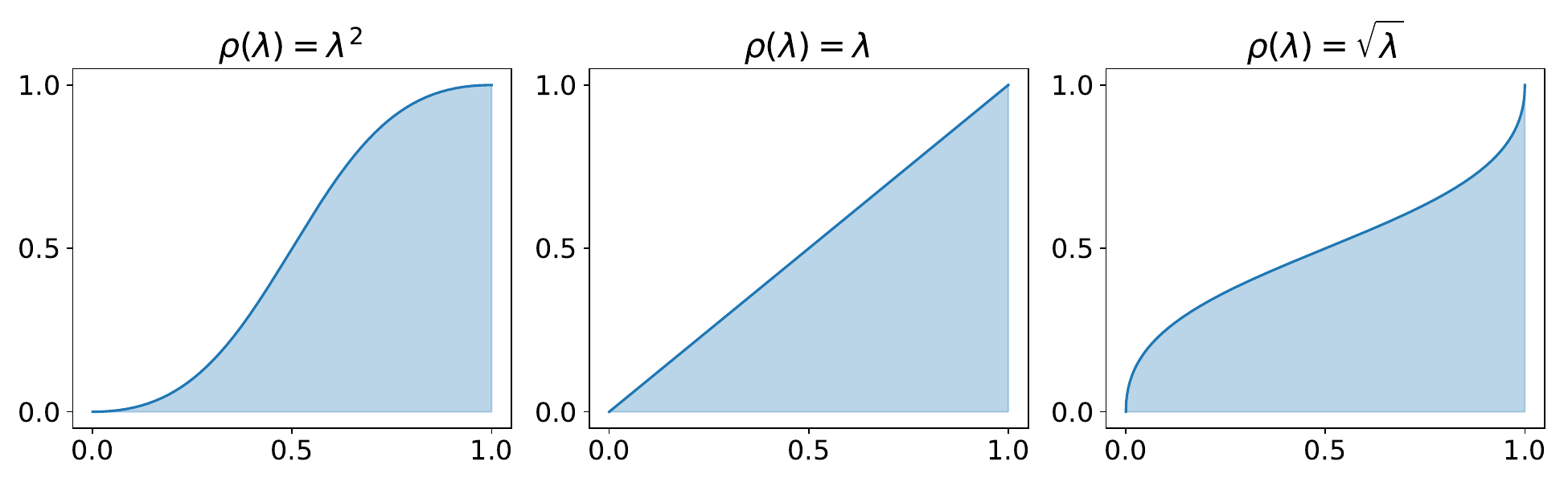}
\caption{Shape of mixing function $\phi$ for different $\rho$.}
\label{fig:shape_visualization}
\end{figure}

Note that the reparameterization form (\ref{eq:representation}) of $\mixingfunc$ produces a regularization effect given certain choices of $\rho$. Specifically, Fig.~\ref{fig:shape_visualization} shows the shapes of the mixing function for various $\rho$ satisfying Theorem~\ref{thm:representation}. Convex $\rho$ flattens $\phi$ near the endpoints, reflecting the principle behind common regularization methods, such as dropout, that small changes in the input preserve the output. Conversely, concave $\rho$ flattens $\phi$ near the midpoint, which reduces the model's dependence on noisy data by forcing both small and large mixing to produce similar losses. As oppose to manually tuning a global mixing distribution that is optimal on average for the whole dataset, \learnablelossmixup automatically learns the optimal $\rho$, and thus $\phi$, for each mixed input, combining the advantages of both regularization regimes.

Importantly, Theorem~\ref{thm:representation} permits flexible neural parameterization of the mixing function $\mixingfunc$. Below, we introduce an instance of such parameterization allowing efficient learning of the mixing function.


\subsection{Neural parameterization of the mixing function}
 We propose a computational efficent parameterization of $\mixingfunc$ by conditioning on the virtual noisy samples. In particular, given a virtual samples $\mixup{\noisy}$ as in (\ref{eq:mixup-sample}), assume our speech enhancement network produces an embedding $\embed{\mixup{\noisy}}$, we use
    \loadeq{eq:embed-mlp}
to parameterize $\mixingfunc$, where $C > 1$, $\sigma$ is the sigmoid function, and MLP is a Multilayer Perception. An example of speech enhancement networks producing such embedding $\embed{\mixup{\noisy}}$ is the UNet described in Table~\ref{table:unet}.

The function in (\ref{eqn:embed_mlp}) satisfies Theorem~\ref{thm:representation} and enables computational efficiency via backpropagation as $\mixingfunc$ in this case is conditioned on an embedding of the mixed samples. Furthermore, when $C>1$, $\rho$ is Eq.~(\ref{eqn:embed_mlp}) is either convex or concave and thus, following the insight in the previous subsection, regularizes the network to prevent overfitting.


\section{Experimental results}\label{sec:experiment}
    In this section, we empirically evaluate and benchmark \learnablelossmixup against state-of-the-art speech enhancement methods on a public speech enhancement dataset.
    
\subsection{Dataset}
    The VCTK dataset \cite{valentini2016speech} is a popular benchmark for single-channel speech enhancement \cite{kim2020t,abdulbaqi2020residual,soni2018time,yao2019coarse,bulut2020low, tran2020}. It contains 30 speakers (28 training, 2 test), 15 noise types (10 training, 5 test), and 8 signal-to-noise ratios (4 training, 4 test). We randomly split 1\% of the training set for validation, and evaluate the models on the test set. The noise types and speakers in the test set are different from those in the training data.

\subsection{Objective metrics}
    We use four popular objective measures for speech enhancement to evaluate the enhanced speech: CSIG, CBAK, COVL, and PESQ. They are computed by comparing the enhanced speech with the corresponding clean reference of each test sample. The first three metrics predict the Mean Opinion Score (MOS) that would result from human perceptual trials: CSIG predicts the signal distortion MOS; CBAK produces the MOS of background-noise intrusiveness; and COVL computes the MOS of the overall signal quality. They produce MOS values from $1$ to $5$. The last metric, PESQ standing for Perceptual Evaluation of Speech Quality, is a broadly used objective measure for speech quality. Its score is in the range between -0.5 to 4.5. For all of the metrics, the higher value corresponds to the better quality of the enhanced speech. In this work, we focus on PESQ as it is most perceptually consistent with our listening test.
    
    
\subsection{Experimental setup}
    
    We predict the log power spectrum of clean speech from that of the corresponding noisy signal and combine the prediction with the noisy phase to produce the enhanced speech. To obtain the spectrograms, we downsample the data to 16kHz, and apply a 512-point STFT with a Hanning window of size 512 and hop length 256. The last frequency bin is removed to form spectrogram inputs of size 64 x 256 x 1. We train a UNet on the log-spectral distance (LSD) loss \cite{tran2020} using the Adam optimizer with learning rate 10\textsuperscript{-4}, $\beta_1=0.5$, $\beta_2=0.9$, for 450 epochs. In addition, we use batch size 64 and apply 0.1 $\ell_2$-regularization. The UNet, summarized in Table~\ref{table:unet}, has $56$ layers with $20,322,818$ parameters in total. Each decoder layer operates on a concatenation of the preceding layer and a skip connection from its corresponding layer in the encoder. Every convolution layer in the UNet, except the last, is followed by batch normalization, dropout with 0.1 drop rate, and Leaky ReLU with 0.2 negative slope. For $\rho$ in (\ref{eqn:embed_mlp}), we set $C=5$, the UNet's bottleneck layer as the embedding, and an MLP with one hidden layer of width 512.
    
    \loadtable{mixup-unet}
    
    For mixup methods, the virtual noisy samples are generated during the training process only. To compute the metrics on the test set, for each test sample ($\noisy$, $\speech$), we compare the prediction $\network(\noisy)$ against the target speech $\speech$. The final metric on the test set is the average of the metrics of the test samples.
    
\subsection{Results}
    
    We benchmarked our proposed framework against classical and state-of-the-art speech enhancement methods including MMSE-GAN \cite{soni2018time}, D+M \cite{yao2019coarse}, UNet \cite{bulut2020low}, Affinity Minimization \cite{tran2020}, T-GSA \cite{kim2020t}, DEMUCS \cite{defossez2020real}, and RHR-Net \cite{abdulbaqi2020residual}. Table~\ref{tab:benchmark-results} shows the numerical results of the aforementioned objective metrics on the test dataset for all benchmarked frameworks. For reference, we also report the objective measures computed for the noisy test signals. The results indicated that \learnablelossmixup outperformed the benchmarked methods in all perceptual metrics excepts CBAK. In particular, it surpassed the state-of-the-art by at least $0.06$ in PESQ.
    \loadtable{benchmark-results}
    \loadtable{ablation-results}

    We further conducted an ablation study on VCTK over 5 random seeds, and list our findings in Table~\ref{table:ablation-results}. Here, learnable label mixup refers to label mixup in which the mixup parameter $\mixupparam$ in (\ref{eq:label-mixup-loss}) is substituted by the mixup function $\mixingfunc$ parameterized by (\ref{eqn:embed_mlp}). Using ERM as a baseline, we found that \learnablelossmixup, favoring VRM, boosted the performance of our UNet model significantly from 3.18 to 3.26 PESQ. Furthermore, consistent with our earlier discussions, label mixup, despite adopting VRM, is inappropriate for speech enhancement: Learnable label mixup fared significantly worse than ERM at 3.10 PESQ. Finally, \lossmixup surpassed ERM slightly by 0.02 points at 3.20 PESQ, demonstrating the benefits of \lossmixup even when the mixing is non-optimal.



\section{Conclusions}\label{sec:conclusions}


We presented \textit{loss mixup}, a training framework resembling data augmentation and Vicinal Risk Minimization, to improve the generalization of deep supervised learning for speech enhancement. Furthermore, we derived neural parameterization of the mixing function, resulting in \textit{learnable loss mixup}, to efficiently learn the optimal mixing function via gradient descent. Experimental results showed that our framework outperforms state-of-the-art speech enhancement methods in the literature. Our results suggested that \lossmixup is a promising approach to boost the generalization of supervised learning for regression problems.

\bibliographystyle{IEEEtran}

\bibliography{mybib}

\begin{thebibliography}{10}
\providecommand{\url}[1]{#1}
\csname url@samestyle\endcsname
\providecommand{\newblock}{\relax}
\providecommand{\bibinfo}[2]{#2}
\providecommand{\BIBentrySTDinterwordspacing}{\spaceskip=0pt\relax}
\providecommand{\BIBentryALTinterwordstretchfactor}{4}
\providecommand{\BIBentryALTinterwordspacing}{\spaceskip=\fontdimen2\font plus
\BIBentryALTinterwordstretchfactor\fontdimen3\font minus \fontdimen4\font\relax}
\providecommand{\BIBforeignlanguage}[2]{{%
\expandafter\ifx\csname l@#1\endcsname\relax
\typeout{** WARNING: IEEEtran.bst: No hyphenation pattern has been}%
\typeout{** loaded for the language `#1'. Using the pattern for}%
\typeout{** the default language instead.}%
\else
\language=\csname l@#1\endcsname
\fi
#2}}
\providecommand{\BIBdecl}{\relax}
\BIBdecl

\bibitem{Boll1979}
S.~F. Boll, ``Suppression of acoustic noise in speech using spectral subtraction,'' \emph{IEEE Transactions on Acoustics, Speech, and Signal Processing}, vol.~27, no.~2, pp. 113--120, 1929.

\bibitem{Lim1978}
J.~S. Lim and A.~V. Oppenheim, ``All-pole modeling of degraded speech,'' \emph{IEEE Trans. on Acoustics, Speech, and Signal Processing}, vol.~26, no.~3, pp. 197--210, 1978.

\bibitem{Lim1979}
------, ``Enhancement and bandwidth compression of noisy speech,'' \emph{Proceedings of the IEEE}, vol.~67, no.~12, pp. 1586--1604, 1979.

\bibitem{Malah1985}
Y.~Ephraimand and D.~Malah, ``Speech enhancement using a minimum mean-square error log-spectral amplitude estimator,'' \emph{IEEE Transactions on Acoustics, Speech, and Signal Processing}, vol.~33, no.~2, pp. 443--445, 1985.

\bibitem{Pascual2017}
S.~Pascual, A.~Bonafonte, and J.~Serra, ``Segan: Speech enhancement generative adversarial network,'' \emph{Interspeech}, 2017.

\bibitem{Williamson2017}
D.~S. Williamson and D.~Wang, ``Time-frequency masking in the complex domain for speech dereverberation and denoising,'' \emph{IEEE/ACM Transactions on Audio, Speech, and Language Processing}, vol.~25, no.~7, 2017.

\bibitem{Rethage2018}
D.~Rethage, J.~Pons, and X.~Serra, ``A wavenet for speech denoising,'' \emph{ICASSP}, 2018.

\bibitem{Germain2018}
F.~G. Germain, Q.~Chen, and V.~Koltun, ``Speech denoising with deep feature losses,'' \emph{Arxiv}, 2018.

\bibitem{Soni2018}
M.~H. Soni, N.~Shah, and H.~A. Patil, ``Time-frequency masking- based speech enhancement using generative adversarial network,'' \emph{ICASSP}, 2018.

\bibitem{Tan2018}
K.~Tan and D.~Wang, ``A convolutional recurrent neural network for real-time speech enhancement,'' \emph{Interspeech}, 2018.

\bibitem{Bulut2020}
A.~E. Bulut and K.~Koishida, ``Low-latency single channel speech enhancement using u-net convolutional neural networks,'' \emph{ICASSP}, 2020.

\bibitem{Mohri2012}
M.~Mohri, A.~Rostamizadeh, and A.~Talwalkar, \emph{Foundations of Machine Learning}.\hskip 1em plus 0.5em minus 0.4em\relax The MIT Press: The MIT Press, 2012.

\bibitem{chapelle2000}
O.~Chapelle, J.~Weston, L.~Bottou, and V.~Vapnik, ``Vicinal risk minimization,'' in \emph{NIPS}, 2020.

\bibitem{tanaka2018joint}
D.~Tanaka, D.~Ikami, T.~Yamasaki, and K.~Aizawa, ``Joint optimization framework for learning with noisy labels,'' in \emph{Proceedings of the IEEE Conference on Computer Vision and Pattern Recognition}, 2018, pp. 5552--5560.

\bibitem{valentini2016speech}
C.~Valentini-Botinhao, X.~Wang, S.~Takaki, and J.~Yamagishi, ``Speech enhancement for a noise-robust text-to-speech synthesis system using deep recurrent neural networks.'' 2016.

\bibitem{zhang2017mixup}
H.~Zhang, M.~Cisse, Y.~N. Dauphin, and D.~Lopez-Paz, ``mixup: Beyond empirical risk minimization,'' \emph{arXiv preprint arXiv:1710.09412}, 2017.

\bibitem{szegedy2016rethinking}
C.~Szegedy, V.~Vanhoucke, S.~Ioffe, J.~Shlens, and Z.~Wojna, ``Rethinking the inception architecture for computer vision,'' in \emph{Proceedings of the IEEE conference on computer vision and pattern recognition}, 2016, pp. 2818--2826.

\bibitem{thulasidasan2019mixup}
S.~Thulasidasan, G.~Chennupati, J.~A. Bilmes, T.~Bhattacharya, and S.~Michalak, ``On mixup training: Improved calibration and predictive uncertainty for deep neural networks,'' in \emph{Advances in Neural Information Processing Systems}, 2019, pp. 13\,888--13\,899.

\bibitem{guo2019mixup}
H.~Guo, Y.~Mao, and R.~Zhang, ``Mixup as locally linear out-of-manifold regularization,'' in \emph{Proceedings of the AAAI Conference on Artificial Intelligence}, vol.~33, 2019, pp. 3714--3722.

\bibitem{verma2019manifold}
V.~Verma, A.~Lamb, C.~Beckham, A.~Najafi, I.~Mitliagkas, D.~Lopez-Paz, and Y.~Bengio, ``Manifold mixup: Better representations by interpolating hidden states,'' in \emph{International Conference on Machine Learning}, 2019, pp. 6438--6447.

\bibitem{kim2020t}
J.~Kim, M.~El-Khamy, and J.~Lee, ``T-gsa: Transformer with gaussian-weighted self-attention for speech enhancement,'' in \emph{ICASSP 2020-2020 IEEE International Conference on Acoustics, Speech and Signal Processing (ICASSP)}.\hskip 1em plus 0.5em minus 0.4em\relax IEEE, 2020, pp. 6649--6653.

\bibitem{abdulbaqi2020residual}
J.~Abdulbaqi, Y.~Gu, S.~Chen, and I.~Marsic, ``Residual recurrent neural network for speech enhancement,'' in \emph{ICASSP 2020-2020 IEEE International Conference on Acoustics, Speech and Signal Processing (ICASSP)}.\hskip 1em plus 0.5em minus 0.4em\relax IEEE, 2020, pp. 6659--6663.

\bibitem{soni2018time}
M.~H. Soni, N.~Shah, and H.~A. Patil, ``Time-frequency masking-based speech enhancement using generative adversarial network,'' in \emph{2018 IEEE International Conference on Acoustics, Speech and Signal Processing (ICASSP)}.\hskip 1em plus 0.5em minus 0.4em\relax IEEE, 2018, pp. 5039--5043.

\bibitem{yao2019coarse}
J.~Yao and A.~Al-Dahle, ``Coarse-to-fine optimization for speech enhancement,'' \emph{arXiv preprint arXiv:1908.08044}, 2019.

\bibitem{bulut2020low}
A.~E. Bulut and K.~Koishida, ``Low-latency single channel speech enhancement using u-net convolutional neural networks,'' in \emph{ICASSP 2020-2020 IEEE International Conference on Acoustics, Speech and Signal Processing (ICASSP)}.\hskip 1em plus 0.5em minus 0.4em\relax IEEE, 2020, pp. 6214--6218.

\bibitem{tran2020}
D.~Tran and K.~Koishida, ``Single-channel speech enhancement by subspace affinity minimization,'' in \emph{INTERSPEECH 2020}.\hskip 1em plus 0.5em minus 0.4em\relax IEEE, 2020.

\bibitem{defossez2020real}
A.~Defossez, G.~Synnaeve, and Y.~Adi, ``Real time speech enhancement in the waveform domain,'' \emph{arXiv preprint arXiv:2006.12847}, 2020.

\end{thebibliography}


\end{document}